\documentclass[12pt]{article}
\input epsf
\begin{document}

\title{\bf Theories of Everything and Hawking's \\
Wave Function of the Universe\thanks{To appear in \sl{The Future 
of Theoretical Physics and Cosmology: Stephen
Hawking 60th Birthday Symposium}, Cambridge University Press with
two additional figures.}}

\author{\bf James B.~Hartle\thanks{hartle@physics.ucsb.edu}\\
{\sl Department of Physics, University of California}\\
{\sl Santa Barbara, CA 93106-9530 USA}}

\date{}
\maketitle

\section{Introduction}
It is an honor, of course,  to participate in this celebration of 
Stephen's 60th birthday and to address such a distinguished audience. 
But for me it is a special pleasure. I first met Stephen over 
thirty years ago when we were both beginning in science at the then
Institute of Theoretical Astronomy here in Cambridge. I have been
following his inspiration ever since and have had the privilege of
working with him on several directions that he has pioneered. 
I want to talk to you about one of these today.

\section{Different Things Fall with the Same\\
Acceleration in a Gravitational Field}
\begin{figure}[t!]
\centerline{\epsfxsize=5in \epsfbox{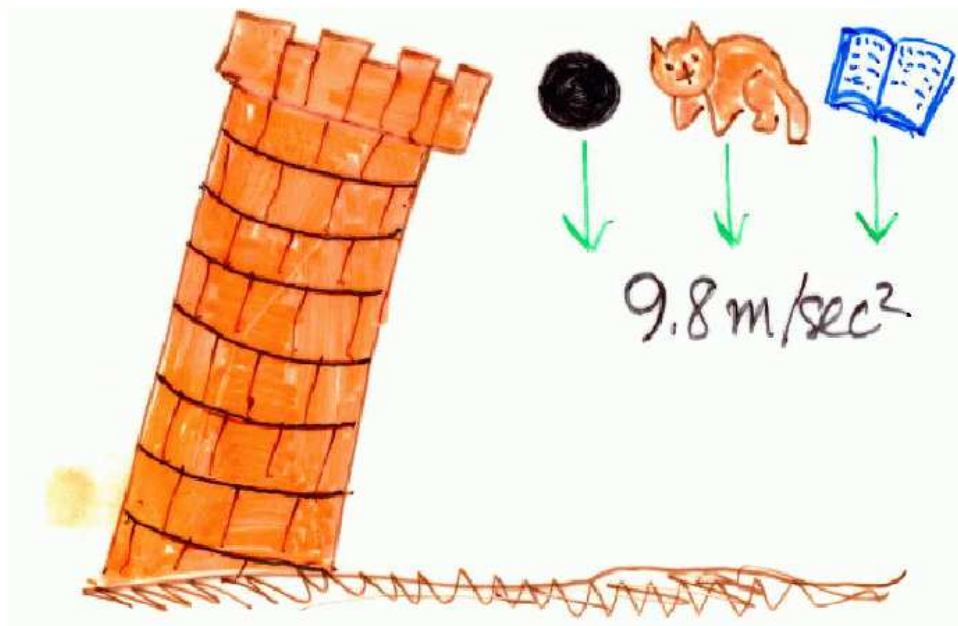}}
\caption{\sl The lecture in a nutshell: A cat, a cannonball, and an
economics textbook all fall to the ground with exactly the same
acceleration if acted upon only by gravity.  The equality of
accelerations of different things in a gravitational field is an
example of a fundamental law of physics.  But that law does not
tell you much about cats, cannonballs, or economics. [From a
transparency used in the lecture].}
\label{trans1}
\end{figure}

If a cat, a cannonball, and an economics textbook are all 
dropped from the same height, they fall to the ground with 
exactly the same acceleration (9.8m/s$^2$) under the
influence of gravity. This equality of the
gravitational accelerations of different things is one of the
most accurately tested laws of physics.  The accelerations are
known to be equal to an accuracy of a few parts in a thousand
billion.  That law, however, tells you very little about cats,
cannonballs, or economics.  That, in a nutshell, is the theme of
my talk today (Figure 1).

Were a cat,  a cannonball, and an economics text
dropped from a real leaning tower, the accelerations would not
be equal to an accuracy of a thousand billion.  Air resistance,
magnetic fields, and a host of other effects mask the purely
gravitational one.  Rather, the equality of accelerations is
revealed to that accuracy only in delicate laboratory
experiments shielded from such extraneous effects, or in the
study of the accelerations of astronomical bodies for which they
are not important. The accuracy record holder at the moment is
the lunar laser ranging demonstration that the Earth and the Moon
fall with the same acceleration toward the Sun \cite{Dic94}.

The manned and unmanned missions to the Moon in the late '60s
and early '70s left behind arrays of corner reflectors. 
A corner reflector is an arrangement of three mirrors making
up the sides of a cube that meet in one corner. It has the
property that any light ray incident on the reflector bounces
from mirror to mirror and is reflected back in exactly the
direction from whence it came.  Since 1969 a systematic program
to monitor the Moon's orbit using these reflectors has been
carried out by the McDonald Observatory in Texas and the
Observatoire de C\^ ote d'Azur in Grasse, France \cite{Dic94}.
High powered lasers send light pulses to the Moon that bounce
off the reflectors. By  measuring precisely the time it takes for
the light pulse to return (about 1 sec), and dividing that by
the velocity of light, the distance to the Moon can be
determined very accurately.  About a billion, billion photons
are sent out ten times a second and one returning photon is
detected every few seconds. The returning photons have been
detected for over 30 years, and as a consequence of this effort,
we know the position of the Moon relative to the Earth to an
accuracy of about a few centimeters.  Study of its orbit shows
it is falling towards the Sun with the same acceleration as the
Earth to an accuracy of a few parts in a thousand billion. 

This law of equality of accelerations of different things is a
cornerstone of Einstein's general relativity --- Stephen's
subject. However, it is not general relativity specifically that
I want to talk to you about today, but about such laws of
physics more generally.

\section{The Fundamental Laws of Physics}

The equality of gravitational accelerations of different things
is an example of a regularity of nature. Everything falls in
exactly the same way. The regularity is universal. No
exceptions! There are other regularities that may apply to
specific systems in similar circumstances.  For
example, if we drop three cats, they might all go `meeeeow' in
roughly the same way on the way down.  Or perhaps there are only
statistical regularities for cats --- 8 out of 10 cats will go
`meeeeow' in the average drop. Other systems will do something
different according to their own particular regularities.
Cannonballs for example do nothing on the way down. But the
equality of accelerations of different things is a special kind
of law because it applies {\it universally} to anything falling
in a gravitational field. It is an example of a fundamental law
of physics.

Identifying and explaining the regularities of nature is the
goal of science. Physics, like other sciences, is concerned with
the regularities exhibited by particular systems.  Stars, atoms,
fluid flows, high temperature super-conductors, black holes, and
the elementary particles are just some of the many examples. 
Studies of these specific systems define the various subfields of
physics --- astrophysics, atomic physics, fluid mechanics, and
so forth. But beyond the regularities exhibited by specific
systems, physics has a special charge.  This is to find the laws
that govern the regularities that are exhibited by {\it all}
physical systems --- without exception, without qualification,
and without approximation.  The equality of gravitational
accelerations of different things is an example.  These are
usually called the {\it fundamental} laws of physics.  Taken
together they are called informally a ``theory of everything''.
Stephen has been a leader in the quest for these universal 
laws \cite{Haw84}. 
Today, I
would like to ask the question: ``How much do we know about the world
if we have a theory of everything?''

\begin{figure}[t]
\centerline{\epsfxsize=5.5in \epsfbox[28 221 585 572]{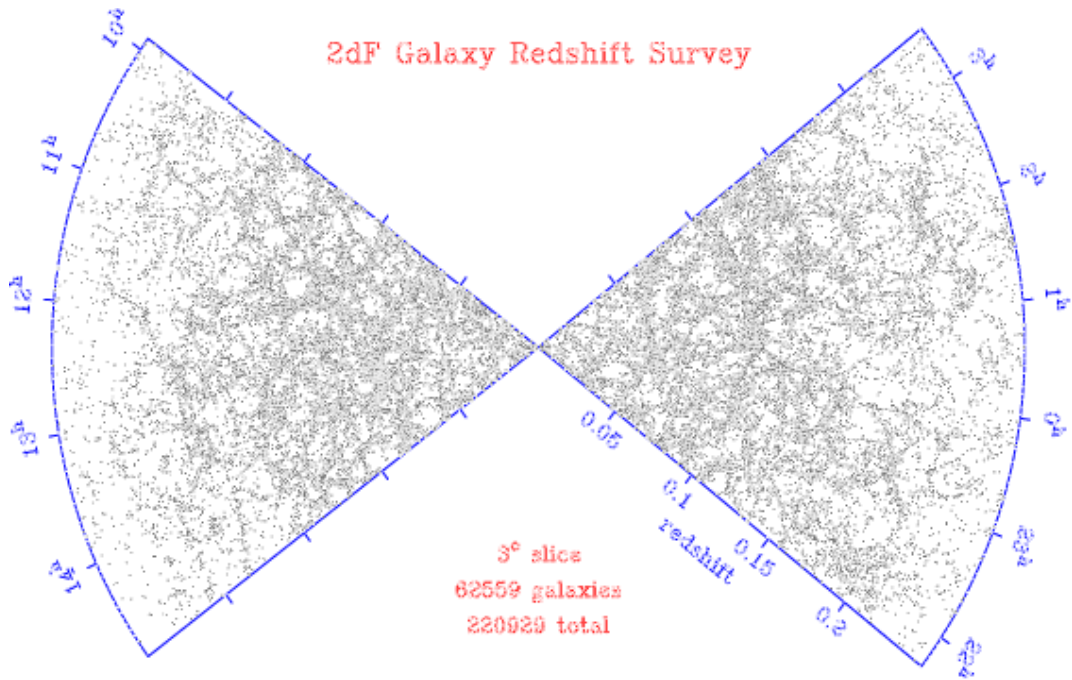}}
\caption{\sl This plot from the 2dF Galaxy Redshift Survey
\cite{CDMS01} shows the location of 62,559 distant galaxies in
two angular slices of the universe approximately 90$^\circ$ wide
and 3$^\circ$ thick as viewed from the Earth. A galaxy's
redshift is used as a measure of its distance from our location
at the center.  The statistics of the distribution of knots,
filaments, and voids in this picture is an example of a large
scale regularity of the universe that is directly linked to its
initial condition. [J.~Peacock and the 2dF Galaxy Redshift
Survey Team]}
\label{2dF}
\end{figure}

Ideas for the nature of the fundamental laws have changed as
experiment and observation have revealed new realms of phenomena
and reached new levels of precision.  But since they have been
studied, it has been thought that the fundamental laws consist
of two parts

\begin{itemize}
\item {\sl Dynamical laws} that govern regularities in {\it time}. 
Newton's 
laws of motion governing the orderly progression of the planets or the 
trajectory of a tennis ball are examples, as is the law that
different things fall with the same acceleration in a gravitational
field and the Einstein equation 
governing the evolution of the universe.

\item {\sl Initial conditions} that govern  
how things started out and therefore most often specify regularities
in {\it space}.  The statistical regularities of the large scale
distribution of the galaxies in the universe is a possible example 
(Figure \ref{2dF}).
\end{itemize}

Laplace is a name associated with this view of a theory of
everything.  He wrote, famously, \cite{Lap51}

\begin{figure}[t]
\centerline{\epsfysize=3in \epsfbox{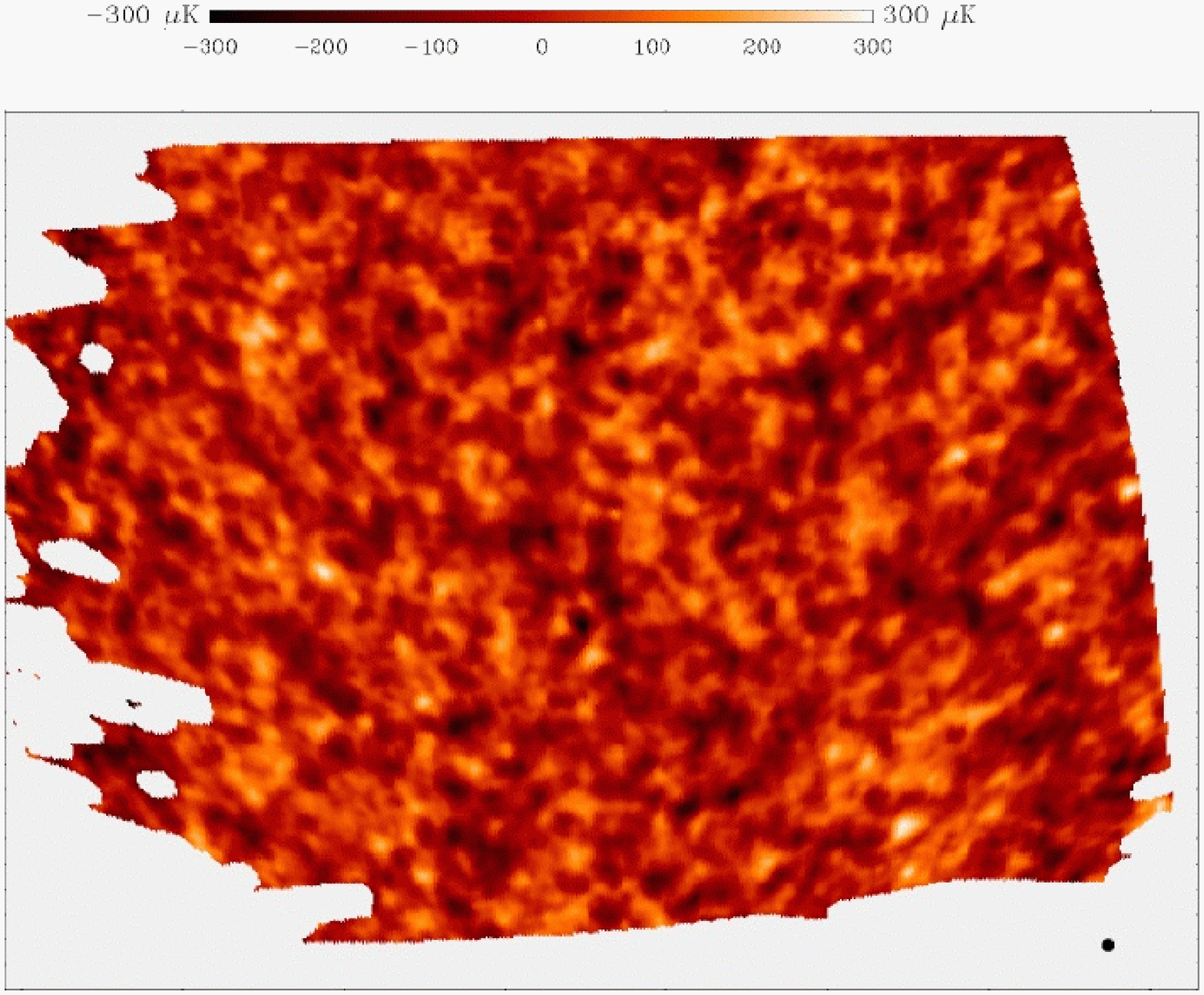}}
\caption{\sl This picture from the Boomerang experiment \cite{Ber00}
shows the temperature distribution across a patch of sky of the
cosmic background radiation. The light and dark regions differ in
temperature by only a few ten millionths of a degree about a mean of
2.73 K.  This is as close as one can come to a picture of the early
universe some hundreds of thousands of years after the big bang. The
temperature variations are evidence for tiny concentrations of
density then which in the intervening 15 billion years grew by the 
action of gravitational attraction  
to become the galaxies we see today with the  large scale distribution
illustrated in  Figure 2.}
\label{boom}
\end{figure}
\begin{quote}
{\sl An intelligence
knowing, at any given instant of time, all forces acting in nature,
as well as the momentary positions of all things of which the
universe consists, would be able to comprehend the motions of the
largest bodies of the world and those of the smallest atoms in one
single formula. 
\dots To it, nothing would be uncertain, both future and past would be 
present before its eyes.}
\end{quote} 
That was the theory of everything circa 1820 --- Newtonian determinism.

Both parts of a theory of everything  are needed to make 
any predictions. Newton's dynamical laws by themselves don't predict the
trajectory of a tennis ball you might throw.  To predict where it goes, you must
also specify the position from which you throw it, the direction, and how fast.
In short, you must specify the ball's {\it initial condition}.  One of Stephen's
most famous achievements is just such an initial condition \cite{Haw84}, but not
for tennis balls.  Stephen's no-boundary initial condition is for the whole
universe.
\begin{figure}[t!]
\centerline{\epsfysize=3.5in \epsfbox{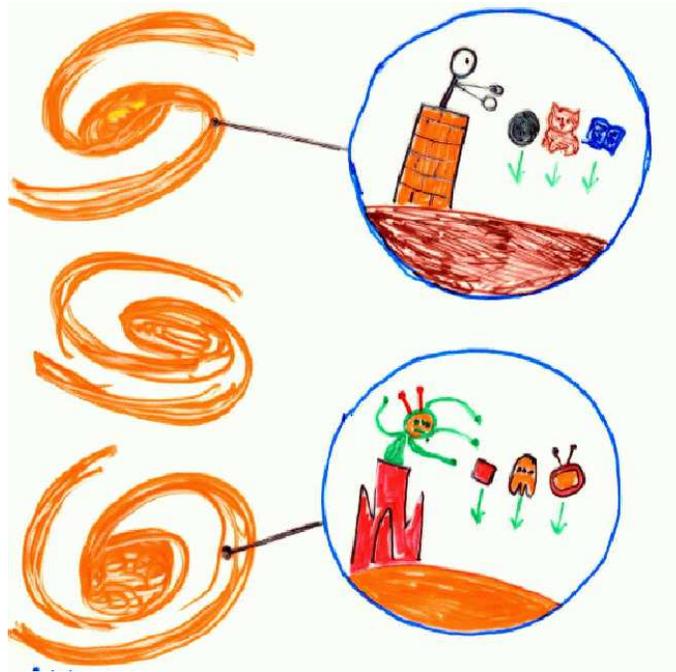}}
\caption{\sl The equality of accelerations of different things in a gravitational
field is an example of a {\rm local} fundamental dynamical law. A body's
acceleration depends only on the gravitational field at its location and not,
say, on the arrangement of distant matter or events in the past that could not be
summarized in this way. This fundamental law can therefore be discovered by
experiments done locally, by experimenters in their laboratories across the
universe. [From a transparency used in the lecture.] }
\label{trans3}
\end{figure}

The search for a theory of the dynamical laws has been seriously under way 
since the time of Newton. Classical mechanics, Newtonian gravity, 
Maxwell's electrodynamics, special and general relativity, quantum 
mechanics, quantum field theory, and superstring theory are but some of 
the milestones in this search. But the search for a theory of the initial 
condition of the universe has
been seriously under way for only the twenty years since Stephen's pioneering
work on the subject.  Why this difference? The examples used above to discuss
regularities governed by the two parts of a theory of everything hint at the
answer.  The trajectory of a tennis ball was used to illustrate the regularities
of dynamical laws, and the large scale distribution of galaxies was used to
illustrate the regularities implied by the law of the initial condition.  There
is a difference in the {\it kind} and {\it scale} of regularities that the two
laws predict.

Dynamical laws predict regularities in time.  It is a fortunate empirical fact
that the fundamental dynamical laws are local --- both in space and time. The trajectory of a
tennis ball depends only on conditions that are nearby both in space and 
time, and not, for example, either on what is  going on in distant parts of 
the universe or a long time ago. This is  
fortunate because that means that dynamical laws can be discovered and
studied in laboratories on Earth and extrapolated, assuming locality, to 
the rest of the universe. For example, because it is local, the law that 
different things fall with the same acceleration in a gravitational field 
can be discovered
by experiments in laboratories here, and indeed all over the galaxy (Figure 4).
Without that simplicity of the dynamical laws in the here and now it is 
possible that we would never have discovered them. 

By contrast, the regularities governed by the law of the initial condition of the
universe occur on large, cosmological scales. The universe isn't simple on small
spatial scales. Look at the disorder or complexity in the room we're in right now
for example. But the universe {\it is} simple on large, cosmological scales ---
more or less the same in one direction as in any other, more or less the same in
one place as in any other. 

The mottled map of the cosmic background radiation in Figure 3 illustrates 
this.  In the beginning the universe was opaque, hot, and glowing. As it 
expanded it got cooler. A few hundred thousand years after the big bang 
the matter cooled to a temperature at which it was transparent.  Light 
from that time has been
propagating to us ever since, cooled to a temperature now of only a few 
degrees above absolute zero.  This is the cosmic background radiation 
and a map of the temperature of the radiation in different directions, 
such as that in Figure 3, is the earliest picture we have of the universe.

The mottled map in Figure 3 may not look like evidence that the early universe was
smooth, but the temperature difference between the light and dark regions 
is only 30 millionths of a degree.  At resolution of your eye the universe would look
completely uniform at this time.  The reason serious research on the initial
condition at the big bang began only recently is that it is only recently that we
have had cosmological observations like this that are detailed enough to reveal
the regularities it governs.

\section{Quantum Mechanics}

There is another way in which our vision of the fundamental laws and the 
nature of a theory of everything has changed since the times of Newton 
and Laplace.  That is quantum mechanics. We don't yet know the final form 
the fundamental laws will take.  But the inference is inescapable from the 
physics of the last
seventy-five years that they will conform to that subtle framework of 
prediction we call quantum mechanics.

In quantum mechanics, any system --- the universe included --- is described by a
wave function $\Psi$. There is a dynamical law called the Schr\"odinger equation
which governs how the wave function changes in time 
\[
i\hbar \frac{d\Psi(t)}{dt} = H\Psi(t) \qquad ({\rm dynamical\ law)}.
\]
Here the operator $H$, called the Hamiltonian, summarizes the dynamical 
theory. There are different forms of $H$ for Maxwell's
electrodynamics, for a theory of the strong interactions, etc.
Like Newton's laws of motion, the Schr\"odinger equation doesn't make any
predictions by itself, it requires an initial condition.  This is
\[
\Psi (0), \qquad ({\rm initial\ condition}).
\]
When we consider the universe as a quantum mechanical system, this
initial condition is Hawking's wave function of the universe \cite{Haw84}

Probabilities are the key difference between classical and quantum mechanics.
Let's first think about probabilities in classical physics.  If I say that
there is a 60\% chance of hitting an audience member if I toss a ball in this
room, I am not expressing a lack of confidence that its trajectory will 
be governed by the deterministic laws of Newtonian mechanics.  Rather the 
60\%
reflects my ignorance of the exact initial speed I'll impart to the ball, 
of the influence of air on its motion, and perhaps my ability to do an 
accurate calculation. If I practice to control the initial condition when 
it's thrown,
the subsequent evolution of the tennis ball becomes more certain. Probabilities
in classical physics result from ignorance.

But in quantum mechanics probabilities are fundamental and uncertainty
inevitable.  No amount of careful determination of the present state of the
tennis ball will achieve certainty for its trajectory. In quantum 
mechanics there is some
probability that a ball will take {\it any} trajectory as it leaves my hand.
However, in classical situations one trajectory --- the one
obeying Newton's laws --- is much more probable than all the others.  The
determinism of classical physics is an approximation, but an approximation on
which we can rely in many practical circumstances. 

\section{A Theory of Everything is Not a Theory of Everything}

My colleague Murray Gell-Mann used to ask me ``If you know the wave 
function of the universe, why aren't you rich?'' 
A quantum mechanical theory of the Hamiltonian and the initial state
{\it does}
predict probabilities for every event that might happen in the universe.  
In that is one sense in which it is a ``theory of everything''. However, only a few
things are predicted with near certainty.  The vast majority of 
alternatives are predicted with approximately 50\%--50\% probabilities 
giving no useful information. That's the sense in which a  ``theory
of everything'' is {\it not} a theory of everything.   

We hope that the probability is high for the form of the
effective dynamical theories that suffice to predict the outcomes of
experiments done in every laboratory constructed so far. We also hope that a
theory of everything predicts some large scale features of the universe with
near certainty such as its approximate smoothness on large scales, the
statistics of the distribution of galaxies illustrated in Figure 2, the vast
disparity between the age of the universe and the time scales characterizing
the fundamental dynamical theory, and so forth.   

But it is too much to hope that an interesting probability like that for 
the FTSE to go up tomorrow is in this category. Since a
rise tomorrow is an event in the history of the universe, a quantum 
mechanical probability could, in principle, be calculated for it from a 
theory of everything. (Although I suspect that it is well beyond our 
present powers of computation.) But it's likely that, after all that
work, the predicted probability
would be 50\% for an upward tick.  That is why you can't get rich knowing 
the wave function of the universe, and that is why a theory of everything 
isn't a theory of everything.  

But there is a deeper reason why a theory of everything doesn't explain
everything --- it's too short. Everyone with a PC knows that picture files are
typically a lot longer than text files. The files for the undergraduate general
relativity text that I am writing, for example, consists of roughly 
1Mb of text and
100Mb of pictures.  A very rough estimate is that it would take a billion,
billion, billion, billion, billion, billion, billion, billion, billion
($10^{81}$) compact disks (CDs) to describe the visible universe at a reasonably
coarse-grained level of classical description. And that is just at one moment
in time! All the matter in the planets, stars, and galaxies in the visible
universe would not be enough to make this number of CDs.

However, the description of the universe can be compressed because it exhibits
regularities on large scales.  Compression is an idea familiar from
computation.  When a large file exhibits regularities its length can be
compressed.  For instance, a string of 1000 zeros `00000000000000000000\dots
0000' in a file is a kind of regularity.  It can be replaced by a shorter
string `1000 zeros'. The regularities summarized by the laws of nature
similarly permit compression. Rather than saying, `cannonballs fall at 
9.8m/s$^2$, cats fall at 9.8m/s$^2$, economics textbooks fall at
9.8m/s$^2$ etc.', the laws of nature permit us to say `{\it everything} falls at 9.8m/s$^2$
near the surface of the Earth'. That is shorter and therefore more useful.
It is possible that everything we see, every detail of every leaf, every 
event in human history, every thought, is a long string which is
compressible to a 
law implementable by a very short computer program. But there is no 
evidence that the universe exhibits such regularity, and even Laplace did 
not propose such a thing.

The initial condition of a deterministic classical theory that did not 
allow ignorance would have to be as detailed as a present description of 
the universe. It seems likely therefore that, were world governed by
classical laws, it might take so many CDs to write out the law of its 
initial condition that we could never make the CDs to do it.

Let's contrast this situation in classical physics with Stephen's 
no-boundary quantum wave function of the universe.  It's given by the 
following simple formula \cite{Haw84}
\[\fbox{$\displaystyle
\Psi = \int \delta g\delta \phi e^{-I[g,\phi]}$} 
\qquad {\left({\rm The\ no-boundary\ wave }\atop{\rm function
\ of\ the\ universe}\right)}
\]
That's short! --- maybe 45 \LaTeX{} key strokes to write out. Furthermore, 
it's as complete a description of the initial condition as it is 
possible to have in quantum mechanics. It implies uncertainty but contains no ignorance.
Presumably, the basic equations of the fundamental dynamical theory are
similarly short.

You might think this a little misleading because I haven't said what 
$\Psi$ means, what $g$ means, and what $\int$ means, etc. But, even  
including the lengths of the ten or so texts that physics students read 
to understand what it all means, a maximum of 10 or 20 Mb would be
needed to state the law --- easily fitting on one CD. This means the
law is the discoverable and implementable, applicable to every happening in the
universe, but predicting the near certainty of only a {\it few} of its many
regularities.  A complete discoverable theory of everything is only possible in
quantum mechanics where some things are predictable but not everything.

Before I left Santa Barbara to come to Cambridge, my colleague Steve Giddings
asked me what I was going to talk about in this lecture. I said I
would speak on the question ``What do we know if we have a theory of 
everything?'' and answer it ``Not all that much.'' He replied something 
like ``I hope you're
going to say something more hopeful than that!'' But it {\it is} a hopeful
message.  It is only because so little of the complexity of the present
universe is predicted by the fundamental universal laws that we can discover
them.

\section{Reduction}

Where do all the other regularities in the universe come from, those 
particular to specific systems --- those of the behavior of cats as they 
fall, those studied by
the environmental sciences like biology, geology, economics, and psychology?
They are the results of chance events that occur naturally over
the history of a quantum mechanical universe.  As my colleague,
Murray Gell-Mann puts it \cite{Man94}, they are {\it frozen accidents}. 
``Chance events of which particular outcomes have a multiplicity
of long-term consequences,  all related by their common ancestry''.

The regularities of cats probably do depend a little bit on the fundamental
physical laws, for example, an initial condition that is smooth across the
universe, leads to three spatial dimensions, etc.  But the origin of most 
of their regularities can be traced to the chance events of four billion 
years of biological evolution.  Cats behave in similar ways because they 
have a common ancestry and develop in similar environments.  The 
mechanisms which  produce those chance events that led to cats are very 
much dependent on fundamental biochemistry and ultimately atomic physics. 
But the particular outcomes of those chances have little to do with the 
theory of everything. 

Do psychology, economics, biology reduce to physics? The answer is YES,
because everything considered in those subjects must obey the
universal, fundamental
laws of physics. Every one of the subjects of study in these sciences ---
humans, market tables, historical documents, bacteria, cats, etc. --- fall with
the same acceleration in a gravitational field.  The answer is NO, because the
regularities of interest in these subjects are not predicted by the 
universal laws with near certainty {\it even in
principle}. They are frozen quantum accidents that produce emergent
regularities. The answer depends upon what you mean by reduce. 

\section{The Main Points Again}

I have always liked the part of the BBC news called ``The Main Points Again''. Here
are my main points again:

\begin{itemize} 

\item The fundamental laws of physics constituting a `theory of
everything' are those which specify the regularities exhibited by every
physical system, without exception, without qualification, and without
approximation.

\item A theory of everything is not (and cannot be) a theory of everything
in a quantum mechanical universe.

\item If it's short enough to be discoverable, it's too short to predict
everything.

\item The regularities of human history, personal psychology, economics,
biology, geology, etc.~are consistent with the fundamental laws of physics, but
do not follow from them.

\end{itemize}
But remember also, especially on this occasion, that all the beautiful
regularities that we observe in the universe, certain or not, predictable or
not, could be the result of quantum chances following from the fundamental
dynamical theory and Hawking's no-boundary wave function of the universe.

\section*{Acknowledgment}
The author's research was supported in part by the National Science
Foundation under grant PHY00-70895.


\begin{thebibliography}{10}

\bibitem{Dic94} {\it e.g.} J.O.~Dickey, {\it et.~al}, {\it Lunar 
Laser Ranging: A Continuing Legacy of the Apollo Program}, {\sl
Science} {\bf 265}, 482 (1994);
J.D. Anderson and J.G~Williams, {\it Long Range
Tests of the Equivalence Principle}, {\sl Class. Quant. Grav.}, {\bf 18},
2447 (2001).

\bibitem{Haw84} {\it e.g.} S.W.~Hawking, 
{\it The Quantum State of the
Universe}, {\sl Nucl.~Phys.~B} {\bf 239}, 257 (1984).

\bibitem{CDMS01} {\it e.g.} M.~Colless, G.~Dalton, S.~Maddox,
W.~Sutherland, {\it et.~al}, {\it The 2dF Galaxy Redshift Survey:
Spectra and Redshifts}, {\sl MNRAS} {\bf 328}, 1039--1063
(2001).

\bibitem{Lap51} P.S.~Laplace, {\it A Philosophical Essay on
Probabilities}, translated from the 6th French edition, Dover
publications, New York (1951). 

\bibitem{Ber00} P.~de Bernardis, {\it et.~al}, {\it A Flat Universe
from High-Resolution Maps of the Cosmic Microwave Background
Radiation}, {\sl Nature} {\bf 404}, 955 (2000).

\bibitem{Man94} M.~Gell-Mann, {\sl The Quark and the Jaguar}, W.~Freeman,
San Francisco (1994).

\end{thebibliography}
\end{document}